\documentclass[lettersize,journal]{IEEEtran}
\usepackage{amsmath,amsfonts,amssymb}
\usepackage{array}
\usepackage{textcomp}
\usepackage{stfloats}
\usepackage{url}
\usepackage{verbatim}
\usepackage{graphicx}
\usepackage{caption}
\usepackage{subcaption}

\usepackage{cite}
\usepackage{multirow}
\usepackage{subcaption}
\usepackage{enumitem}
\usepackage{algorithm}
\usepackage[noend]{algpseudocode}
\usepackage{tcolorbox}
\usepackage{booktabs}
\usepackage{etoolbox}
\usepackage{bm}
\usepackage{xspace}
\usepackage{url}
\usepackage{textcomp}
\usepackage{xcolor}
\usepackage{cite}
\usepackage{listings}
\usepackage{wrapfig}
\usepackage{tikz}
\def\BibTeX{{\rm B\kern-.05em{\sc i\kern-.025em b}\kern-.08em
    T\kern-.1667em\lower.7ex\hbox{E}\kern-.125emX}}
\usepackage{balance}
\begin{document}
\title{The Prompt Alchemist: Automated LLM-Tailored Prompt Optimization for Test Case Generation}

\markboth{Journal of \LaTeX\ Class Files,~Vol.~18, No.~9, September~2020}%
{The Prompt Alchemist: Automated LLM-Tailored Prompt Optimization for Test Case Generation}

\author{
\IEEEauthorblockN{Shuzheng~Gao$^{1}$, Chaozheng~Wang$^{1}$, Cuiyun~Gao$^{2*}$, Xiaoqian Jiao$^2$, Chun Yong Chong$^3$, \\ Shan Gao$^4$, Michael R. Lyu$^1$}

\IEEEauthorblockA{$^1$ The Chinese University of Hong Kong, Hong Kong, China}

\IEEEauthorblockA{$^2$ Harbin Institute of Technology, Shenzhen, China}

\IEEEauthorblockA{$^3$ School of Information Technology, Monash University Malaysia, Malaysia}

\IEEEauthorblockA{$^4$ Independent Researcher, China}

\IEEEauthorblockA{szgao23@cse.cuhk.edu.hk,
adf111178@gmail.com,
gaocuiyun@hit.edu.cn,
210110210@stu.hit.edu.cn,
chong.chunyong@monash.edu,
gaoshan\_cs@outlook.com,
lyu@cse.cuhk.edu.hk}
\thanks{$^{\ast}$ Corresponding author. The author is also affiliated with Peng Cheng Laboratory.}}

\definecolor{mydarkgreen}{RGB}{0,150,0}
\lstset{
  language=Java,
  basicstyle=\ttfamily\footnotesize,
  keywordstyle=\color{blue}\bfseries, 
  commentstyle=\color{gray}\itshape,
  stringstyle=\color{mydarkgreen},
    breakatwhitespace=false,         
    breaklines=true,                 
    captionpos=b,                    
    keepspaces=true,                 
    numbers=none,                    
    numbersep=5pt,                  
    showspaces=false,                
    showstringspaces=false,
    showtabs=false,                  
    tabsize=2,
}

\renewcommand{\algorithmicrequire}{ \textbf{Input:}}
\renewcommand{\algorithmicensure}{ \textbf{Output:}}

\algnewcommand{\LeftComment}[1]{\Statex \(\triangleright\) #1}
\newcommand{\resetalgocf}{\setcounter{AlgoLine}{0}}
\newcommand*{\circled}[1]{\lower.7ex\hbox{\tikz\draw (0pt, 0pt)%
    circle (.5em) node {\makebox[1em][c]{\small #1}};}}

\newcommand{\gsz}[1]{\textcolor{orange}{{#1}}}
\newcommand{\yun}[1]{\textcolor{blue}{{#1}}}

\newcommand{\tool}{MAPS\xspace}

\maketitle

\begin{abstract}
Test cases are essential for validating the reliability and quality of software applications. Recent studies have demonstrated the capability of Large Language Models (LLMs) to generate useful test cases for given source code. However, the existing work primarily relies on human-written plain prompts, which often leads to suboptimal results {since} the performance of LLMs can be highly influenced by the prompts. Moreover, these approaches use the same prompt for all LLMs, overlooking the fact that different LLMs might be best suited to different prompts. Given the wide variety of possible prompt formulations, automatically discovering the optimal prompt for each LLM presents a significant challenge. Although there are methods on automated prompt optimization in the natural language processing field, they are hard to produce effective prompts for the test case generation task. First, the methods iteratively optimize prompts by simply combining and mutating existing ones without proper guidance, resulting in prompts that lack diversity and tend to repeat the same errors in the generated test cases. Second, the prompts are generally lack of domain contextual knowledge, limiting LLMs' performance in the task.

In this paper, we introduce \tool, an LL\textbf{M}-t\textbf{A}ilored \textbf{P}rompt generation method for te\textbf{S}t case generation. \tool comprises three main modules: Diversity-guided Prompt Generation, Failure-driven Rule Induction, and Domain Contextual Knowledge Extraction. Specifically, in the \textit{Diversity-Guided Prompt Generation} module, \tool creates varied prompts by exploring diverse modification paths during the optimization process. It prevents the optimization process from converging to local optima. The \textit{Failure-driven Rule Induction} module aims at identifying promising optimization direction by reflecting common failures in generated test cases, in which the reflection outputs are softly integrated into prompts based on a rule transformation method. The \textit{Domain Contextual Knowledge Extraction} module aims at enriching the prompts with related domain knowledge by incorporating both in-file and cross-file context information. To evaluate the effectiveness of \tool, we compare it with four state-of-the-art prompt optimization methods across three popular LLMs. The experimental results demonstrate that our method outperforms baseline methods by a large margin, achieving a 6.19\% higher line coverage rate and a 5.03\% higher branch coverage rate on average. Moreover, experiments on different LLMs show that our method can effectively find the most suitable prompt for each LLM.
\end{abstract}

\begin{IEEEkeywords}
Software testing and debugging, Test Case Generation, Large Language Models.
\end{IEEEkeywords}

\section{Introduction}\label{sec:intro}

Test cases play a crucial role in validating the reliability and quality of software applications~\cite{DBLP:conf/icse/RafiMPM12,DBLP:conf/icse/AlmasiHFAB17}. By allowing developers to identify and rectify bugs and defects at the early development stage, it {remarkably}
enhances the overall stability of the software~\cite{DBLP:journals/software/Runeson06}. However, manually writing test cases is a challenging and time-consuming task. Consequently, the task of test case generation, which aims at creating high-quality test cases automatically, has attracted both developers' and researchers’ attention in recent years~\cite{DBLP:conf/oopsla/PachecoE07,tufano2020unit,DBLP:journals/tse/WangHCLWW24}.

Traditional test case generation methods such as Evosuite~\cite{DBLP:conf/sigsoft/FraserA11} and Randoop~\cite{DBLP:conf/oopsla/PachecoE07} mainly employ search-based and constraint-based techniques to craft test suites. Recent advancements in deep learning have introduced many learning-based test generation approaches. For instance, AthenaTest~\cite{tufano2020unit} fine-tunes BART~\cite{DBLP:conf/acl/LewisLGGMLSZ20} on a dataset designed for test generation. A3Test~\cite{DBLP:journals/corr/abs-2302-10352} further incorporates assertion knowledge and a test signature verification mechanism {for achieving}
better results. These models {aim at leveraging}
general programming knowledge acquired from extensive developer-written code corpora to generate more comprehensive and meaningful tests. {Recently,} Large Language Models (LLMs), such as ChatGPT~\cite{ChatGPT}, have
gained widespread adoption in various Software Engineering (SE) tasks, including test case generation, and show promising results. Due to their powerful zero-shot capabilities, LLMs can be directly deployed for downstream tasks through prompt engineering without requiring fine-tuning~\cite{DBLP:conf/nips/KojimaGRMI22}. For example, ChatUniTest~\cite{DBLP:journals/corr/abs-2305-04764} harnesses the capabilities of LLMs and employs a generation-validation-repair mechanism to rectify errors in generated test cases. Yuan et al.~\cite{DBLP:journals/pacmse/Yuan0DW00L24} evaluate ChatGPT's performance in test case generation and enhance it through an iterative test refinement process.

However, the existing {LLM-based} work primarily relies on human-written plain prompts, which often leads to suboptimal results since the performance of LLMs can be highly influenced by the prompts~\cite{DBLP:conf/kbse/GaoWGWZL23,DBLP:conf/emnlp/Gonen0BSZ23}.
Additionally, different LLMs may be best suited to different prompts. For instance, as shown in Table~\ref{tab:prompt}, our preliminary experiments of three 
prompts on a subset of Defects4J~\cite{DBLP:conf/issta/JustJE14} reveals varying performance across different LLMs. Specifically, the best prompt on ChatGPT achieves a 24.46\% line coverage rate, while the worst one achieves only 21.92\%, indicating that prompt choice can {greatly}
influence the performance of LLMs for test case generation and plain prompts may not yield satisfactory results. Furthermore, our analysis reveals that the prompt performing best with ChatGPT actually performs worst when applied to Llama-3.1~\cite{llama3}. Therefore, given the {considerable}
time required for manual prompt design, the automated generation of tailored prompts for different LLMs is worth studying but has not received adequate attention.

\begin{table}[t]
 \centering
 \aboverulesep=0ex
\belowrulesep=0ex
\caption{Comparison of test case generation prompts and their line coverage rates across different LLMs using the Defects4J~\cite{DBLP:conf/issta/JustJE14} benchmark.}
\label{tab:prompt}
\scalebox{0.9}{
\begin{tabular}{l|c|c}
\toprule
\textbf{Prompt} & \textbf{ChatGPT} & \textbf{Llama-3.1}\\ 
\midrule
 Write unit tests for the provided Java classes to  & \multirow{2}{*}{21.92\%} & \multirow{2}{*}{\textbf{26.45\%}} \\
 test the methods and functionalities of each class. & &  \\
\midrule
Write unit tests for the given Java classes to  & \multirow{2}{*}{\textbf{24.46\%}} & \multirow{2}{*}{24.07\%} \\
 ensure proper functionality of the methods. & &  \\
\midrule
Write test cases for the given Java class to  & \multirow{2}{*}{22.90\%} & \multirow{2}{*}{25.80\%} \\
 ensure the correct behavior of its methods.  & &  \\
\bottomrule
\end{tabular}}
\end{table}

To {achieve}
LLM-tailored prompts, one potential approach is to leverage prompt optimization methods from {the} Natural Language Processing (NLP) {field}~\cite{DBLP:conf/iclr/Guo0GLS0L0Y24,DBLP:conf/iclr/ZhouMHPPCB23}. These methods typically use LLMs and evolutionary algorithm~\cite{holland1992adaptation,storn1997differential} to iteratively search the discrete natural language space for effective prompts through a generate-and-validate approach. However, when applied to test case generation, these methods fall short of achieving promising results due to three main limitations:
\textit{(1) Low diversity in generated prompts.} These methods optimize prompts by simply combining and mutating existing ones using LLMs, while ignoring the diversity in generated prompts, {which potentially leads to insufficient exploration of the vast natural language search space}. 
Consequently, the optimization process may converge prematurely to local optima, hindering the discovery of the most suitable prompt.
\textit{(2) Lack of proper guidance {on avoiding common errors}.}
Existing methods generate new prompts based solely on existing ones without {considering the recurring errors.}
As a result, test cases produced by optimized prompts often exhibit the same issues as those generated by unoptimized prompts. Therefore, it is {important}
to effectively guide the optimization process with directed improvement and prevent LLMs from making 
recurring 
errors.
\textit{(3) Absence of domain contextual knowledge.} Existing LLM-based test case generation approaches~\cite{DBLP:journals/pacmse/Yuan0DW00L24,DBLP:journals/pacmse/Ryan0S00RR24} typically utilize only the focal method or limited in-file context information. They lack necessary domain contextual knowledge such as subclass inheritance and class invocation information, which {is}
crucial for generating {accurate}
test cases. Given the complex inheritance and invocation relationships between classes and functions in real-world projects, {it is difficult for LLMs to 
infer such information}.


In this paper, we propose \tool, the first LL\textbf{M}-t\textbf{A}ilored \textbf{P}rompt generation method for te\textbf{S}t case generation. \tool effectively and automatically generates suitable prompts for different LLMs through three key modules: diversity-guided prompt generation, failure-driven rule induction, and domain contextual knowledge extraction.
The \textit{diversity-guided prompt generation} module creates varied prompts by exploring diverse modification paths during prompt optimization. This approach prevents premature convergence to local optima, ensuring a more comprehensive exploration of the prompt space. 
{The \textit{failure-driven rule induction} module aims at identifying promising optimization direction by reflecting common errors in generated test cases and guide the optimization process by transforming the reflection results into rules. These rules are then incorporated {into}
the prompt to prevent recurring errors.} 
Furthermore, the \textit{domain contextual knowledge extraction} module provides LLMs with both in-file and cross-file context information, such as inheritance relationship information, to help them generate accurate test cases. The optimized prompt, induced rules, and extracted context information are then integrated together to form the final prompt for test case generation.
To evaluate the effectiveness of \tool, we conduct experiments on a popular benchmark Defects4J~\cite{DBLP:conf/issta/JustJE14}. We apply \tool to three popular LLMs including ChatGPT~\cite{ChatGPT}, Llama-3.1~\cite{llama3}, and Qwen2~\cite{qwen2} and compare it with four state-of-the-art prompt optimization approaches. The experimental results demonstrate that \tool outperforms baseline methods by a large margin, achieving a 6.19\% higher line coverage rate and a 5.03\% higher branch coverage rate on average. Besides, experiments on different LLMs reveal that \tool can effectively generate the most suitable prompt for each LLMs, surpassing manually designed prompts.

\textbf{Contributions.} In summary, the main contributions of this work are as follows:
\begin{enumerate}
    \item To the best of our knowledge, this paper presents the first study on automatically {producing}
    LLM-tailored prompt for test case generation.
    \item We propose a novel method \tool that effectively improves the prompt optimization process 
by integrating diversity-guided prompt generation, failure-driven rule induction and domain contextual knowledge extraction. 
    \item Extensive experiments on three popular LLMs demonstrate that our method substantially outperforms baseline approaches and effectively generate tailored prompts for different LLMs. 
\end{enumerate}

\textbf{Organization.} The rest of this paper is organized as follows. Section~\ref{sec:back} describes the background ans shows our motivating examples. Section~\ref{sec:approach} details the three components in the proposed \tool, including the domain contextual knowledge extraction, diversity-guided prompt generation and failure-driven rule induction. Section~\ref{sec:setup} describes the evaluation methods, including the research questions, datasets, baselines, and implementation details. Section~\ref{sec:result} presents the experimental results. Section~\ref{sec:discuss} discusses some cases and threats to validity. Section~\ref{sec:conclusion} concludes the paper.
\section{Background and Motivating Example}\label{sec:back}

\subsection{Background}
In this work, we concentrate on black-box LLM-based Automatic Prompt Optimization (APO)~\cite{DBLP:conf/iclr/Guo0GLS0L0Y24,DBLP:journals/corr/abs-2402-02101}, given the widespread adoption and powerful capabilities of black-box LLMs. APO utilizes LLMs to optimize prompts by iteratively searching for the most effective ones within the discrete space of natural language. Formally, for a task, we work with a black-box model $M$, a small development set $D_{dev}$, a test set $D_{test}$, and a scoring function $s(\cdot)$. {APO aims at discovering an advanced prompt $p$ based on $D_{dev}$ from the natural language space that maximizes the performance of $M$ on the test set $D_{test}$. The prompt $p$ is expected to guide the model directly generate high-quality responses instead of time-consuming multi-iteration generation during test time.}
A typical APO framework operates as follows. First, it begins with a set of seed prompts which can be obtained either manually or through automatic techniques. Then the seed prompts are used to generate responses for $D_{dev}$ via $M$ and the responses are evaluated using the scoring function $s(\cdot)$, such as the line coverage rate in test case generation. Prompts that perform well are retained, while those that do not are discarded. Using the retained prompts, the APO methods query $M$ to generate new prompts. For example, a representative method OPRO~\cite{DBLP:conf/iclr/Yang0LLLZC24} generates new prompts by prompting LLMs with the prompt ``Generate an instruction that is different from all the instructions and has a higher score than all the instructions above''. The newly generated prompts will be integrated with the retained prompts for next iteration optimization. After several iterations, the best prompt on $D_{dev}$ will be used as the final optimized prompt for $D_{test}$. 

\subsection{Motivating Examples}
We first conduct a preliminary study by applying {existing APO methods}
to real-world test case generation on Defects4J and find that it struggles to produce well-performing prompts. By analyzing its optimized prompts and generated test cases, we identify three main problems of current APO methods.

\textbf{Observation 1 [Low Diversity of Prompts Generated during Optimization Process]:}
{First, upon inspection of the prompts generated during the optimization process, we find that they tend to exhibit similar phrases and lack diversity.}
{Table~\ref{tab:example1} presents some examples of prompts generated by OPRO~\cite{DBLP:conf/iclr/Yang0LLLZC24} which contain similar phrases such as ``\textit{Create unit tests to}'' and ``\textit{the provided Java classes}''.} {The low diversity constrains the optimization process to a small portion of the discrete natural language search space, limiting exploration of potentially more effective alternative phrases. This makes the search process susceptible to convergence at local optima and yielding suboptimal performance.}
Therefore, to deal with this problem, the first key idea of our method is to improve the diversity of generated prompts by enforcing them to use different modification methods in the optimization process.

\begin{table*}[t]
 \centering
 \aboverulesep=0ex
\belowrulesep=0ex
\caption{{Examples of prompts generated by OPRO~\cite{DBLP:conf/iclr/Yang0LLLZC24}}
in the optimization process. The underlined part represents the similar pattern {among the prompts.}}
\label{tab:example1}
\scalebox{1.1}{
\begin{tabular}{l}
\toprule
\textbf{Prompt} \\
\midrule
1. \underline{Create unit tests to} verify the correctness of method implementations in \underline{the provided Java classes}. \\
2. \underline{Create unit tests to} validate the functionality of specific methods within \underline{the provided Java classes}. \\
3. \underline{Create unit tests to} ensure \underline{that the methods} in the supplied Java classes behave as expected. \\
4. \underline{Create unit tests to} confirm \underline{that the methods} behave as expected and produce the correct results. \\
\bottomrule
\end{tabular}}
\end{table*}


\begin{lstlisting}[caption={One example showing recurring errors made by the seed prompt and optimized prompt. The error lines are highlighted in red.},escapeinside={(*@}{@*)},label={lst:example2}]
// Focal method:
public class TimeSeries extends Series implements Cloneable,Serializable{
  public TimeSeries createCopy(int start, int end)
    throws CloneNotSupportedException {
    if(start < 0){throw new IllegalArgumentException("Requires start >= 0.");}
    if(end < start){throw new IllegalArgumentException("Requires start <= end.");}
    ...
// Test case generated by seed prompt:
public void testCreateCopy_empty() {
  TimeSeries timeSeries = new TimeSeries("Test");
  TimeSeries copy = (*@\colorbox{red!30!}{timeSeries.createCopy(0,}@*)
    (*@\colorbox{red!30!}{timeSeries.getItemCount()-1);}@*)
  ...
// Test case generated by optimized prompt:
public void testCreateCopy_empty() {
  TimeSeries timeSeries = new TimeSeries("EmptyTest");
  TimeSeries copy = (*@\colorbox{red!30!}{timeSeries.createCopy(0,}@*)
    (*@\colorbox{red!30!}{timeSeries.getItemCount()-1);}@*)
  ...
\end{lstlisting}

\textbf{Observation 2 [Recurring Common Errors Across Iterations]:} Additionally, by analyzing the generated test cases on $D_{dev}$ in different iterations, we find that the test cases generated by optimized prompts tend to exhibit the same errors as the unoptimized ones. For example, as shown in Listing~\ref{lst:example2}, both the test cases generated by the seed prompt and the optimized prompt lack exception handling statements and encounter the same runtime errors. Existing prompt optimization methods rely solely on current prompts without proper guidance, making it difficult to achieve directed improvements and address the errors
made by current prompts. To tackle these challenges, we propose to leverage failed test cases to identify shortcomings in current prompts. 
Specifically, we make LLMs reflect common errors in generated test cases and softly incorporate the reflection outputs into prompts as {concise} rules to help LLMs avoid making recurring errors. 

\begin{lstlisting}[caption={One example illustrating the issue of lacking domain context information. The error lines are highlighted in red.},escapeinside={(*@}{@*)},label={lst:example3}]
// Focal method:
public abstract class AbstractCategoryItemRenderer 
    extends AbstractRenderer implements CategoryItemRenderer, 
    Cloneable, PublicCloneable, Serializable {
    public CategoryItemLabelGenerator getItemLabelGenerator
        ...
// Test case:
public void testFindRangeBoundsValidDataset() {
    AbstractCategoryItemRenderer renderer = (*@\colorbox{red!30!}{new AbstractCategoryItemRenderer();}@*)
    ...
\end{lstlisting}

\textbf{Observation 3 [Lack of Domain Contextual Knowledge]:} Finally, we thoroughly analyzed the focal methods where all prompts and LLMs failed to generate correct test cases. The primary issue identified is the lack of domain contextual knowledge. As illustrated in Listing~\ref{lst:example3}, the given focal method is from an abstract class ``\texttt{AbstractCategoryItemRenderer}''. The generated test case directly initialize with an abstract class which leads to the error: ``\textit{\texttt{AbstractCategoryItemRenderer} is abstract; cannot be instantiated}''. Without knowledge of its subclasses, LLMs cannot generate test cases that correctly initialize this class and invoke the method. Therefore, another key idea of \tool is to extract the relevant domain contextual information and provide it to the LLMs {for capturing}
contextual knowledge.

\section{Proposed Approach}\label{sec:approach}

 \begin{figure*}[t]
    \centering
    \includegraphics[width=1\textwidth]{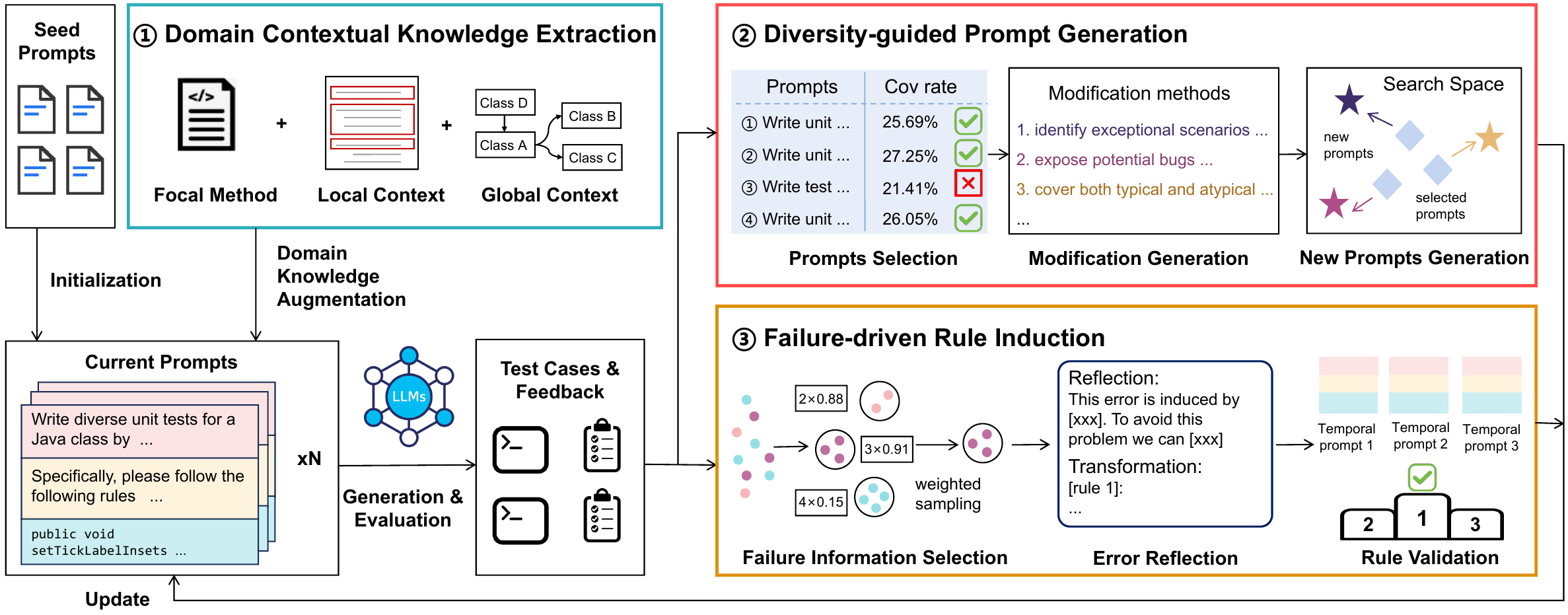}
    \caption{Overview of \tool’s workflow.}
    \label{fig:overview}
\end{figure*}

\subsection{Overview}
\begin{figure*}
    \centering
    \includegraphics[width=0.85\textwidth]{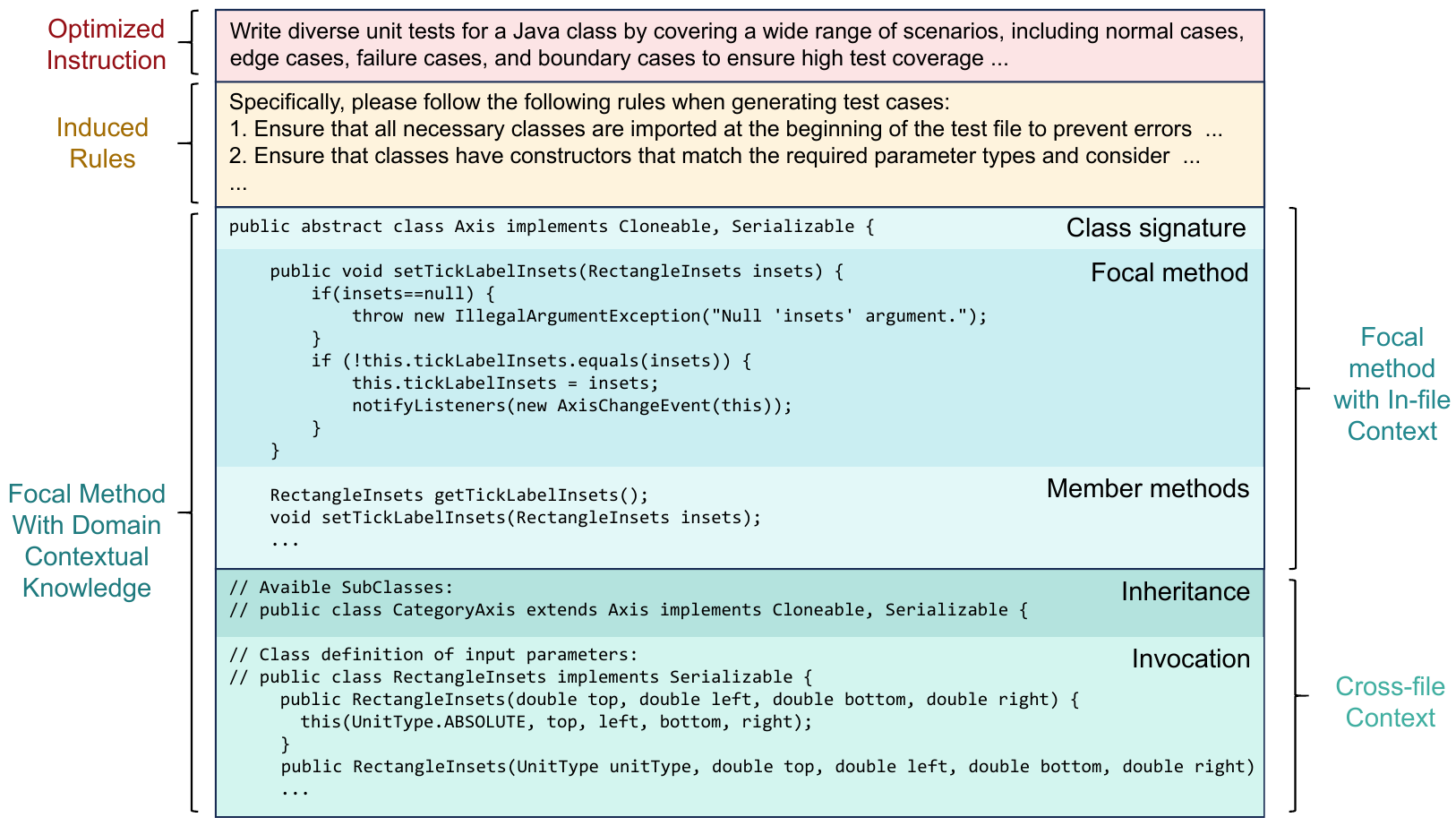}
    \caption{An illustration of the format of final prompt and extracted context information.}
    \label{fig:context}
\end{figure*}

We provide an overview of \tool's workflow in Fig.~\ref{fig:overview}. \tool starts with a set of seed prompts and augments the focal methods with both in-file and cross-file context information through the \circled{1}\textit{domain contextual knowledge extraction} module. In each iteration, \tool first evaluates the performance of the current prompts on the small development set. The \circled{2}\textit{diversity-guided prompt generation} module then selects the top-performing prompts and infers diverse modification methods, which are used to help generate creates varied prompts. In the \circled{3}\textit{failure-driven rule induction} module, \tool aggregates and selects representative
failure information from failed test cases and induces {concise} rules to avoid such failures using a reflection-validation method. As shown in Algorithm~\ref{alg:tool}, this iterative optimization process continues until reaching the maximum iteration number $I$. Finally, the best-optimized instruction from \textit{diversity-guided prompt generation}, induced rules, and extracted context information are integrated to construct the final prompt for test case generation. 
Fig.~\ref{fig:context} illustrates the format of the final prompt. 
In the following sections, we will introduce these three modules in details.

\begin{algorithm}[t]
\caption{Algorithm of \tool}
\label{alg:tool}
\begin{algorithmic}[1]
\Require Iteration number $I$, Seed prompt $P$, Domain contextual knowledge $C$, LLM $M$
\Ensure Final prompt
\State $R\gets\varnothing$,  $H\gets\varnothing$ \Comment{Initialize the set of rules $R$ and handled failures $H$ in previous iteration}
\For{each $i$ in $I$} 
\State Evaluate \textsc{Format($P$, $R$, $C$)} on the sampled development set
\State $P$, $NR$, $NH$ = \textsc{PromptImprovement}($P$, $R$, $H$, $M$, $C$)
\State $R\gets R \cup NR$, $H\gets H \cup NH$
\EndFor
\State $p \gets \textsc{SelectTop}(P, 1)$ \Comment{Select the best prompt from $P$}
\State \textbf{return} \textsc{Format($p$, $R$, $C$)} \Comment{Formalize the final prompt}
\end{algorithmic}
\end{algorithm}


\subsection{Domain Contextual Knowledge Extraction}
The domain contextual knowledge extraction module aims to provide LLMs with {related} project-level context information, enabling them to generate accurate test cases. As illustrated in Fig.~\ref{fig:context}, the contextual knowledge is divided into two categories: \textit{in-file contextual knowledge} and \textit{cross-file contextual knowledge}.

\begin{itemize}
    \item \textbf{In-file Contextual Knowledge} contains the class signature, focal method, and the signatures of member methods. The class signature includes the type and name of the class containing the focal method, which could help LLMs avoid direct initialization of abstract or private classes. The focal method is the specific method {to generate} 
    test cases. Following previous research~\cite{tufano2020unit,DBLP:journals/corr/abs-2302-10352}, we also incorporate the function signatures of other member methods within the class, as the focal method may invoke them, and these signatures can guide the correct usage of these functions.

    \item \textbf{Cross-file Contextual Knowledge} refers to {the} context information from other files within the project. We propose to extract two types of cross-file information {that are} critical for test case generation but ignored in previous work, namely \textit{class inheritance information} and \textit{class invocation information}. For focal methods from abstract or private classes, we scan the entire project to locate their subclasses and extract the class signatures. This subclass information enables LLMs to properly instantiate the class within the test case. Furthermore, for {the} class invocation information, we identify the types of arguments in the focal method, trace the definitions of user-defined types, and extract their signatures and constructors. {This invocation information} 
aids LLMs in using correct input arguments for the focal method. 
\end{itemize}

\subsection{Diversity-guided Prompt Generation}
The diversity-guided prompt generation module aims at producing diverse prompts to foster a more comprehensive exploration of the prompt space by enforcing them to use different modification methods. As illustrated in Fig.~\ref{fig:overview}\circled{2} and Algorithm~\ref{alg:prompt}, after evaluating the performance of current prompts on the evaluation set, \tool selects the top-$K$ prompts with the highest average line coverage and branch coverage. 
Using these selected samples, {\tool first leverages the LLM $M$ to generate $N$ distinct modification methods for the current prompts based on a modification prompt template shown in Fig.~\ref{fig:template}} (a) (Lines 4), where $N=\textsc{Size}(P)-K$ {and $\textsc{Size}(P)$ indicates the number of seed prompts,} 
to maintain a constant {prompt number} 
{following} 
previous work~\cite{DBLP:conf/iclr/Guo0GLS0L0Y24}. These modification methods serve as diverse exploration directions within the discrete natural language search space. \tool then leverages LLM $M$ to generate new prompts based on each modification method sequentially (Lines 5-7). Finally, the selected prompts and the newly generated prompts are combined to serve as the {new} 
prompts for the next iteration of optimization.

\subsection{Failure-driven Rule Induction}
The failure-driven rule induction {module}
aims at identifying promising optimization direction {by avoiding}
LLMs to make recurring errors. It leverages common failures in generated test cases to identify the parts where existing prompts most need improvement and induces rules to optimize the prompt using a reflection-validation method. As shown in Fig.~\ref{fig:overview}\circled{3}, this process contains three phases: failure information selection, error reflection, and rule validation. The details are illustrated in Algorithm~\ref{alg:prompt}.

\begin{algorithm*}[t]
\caption{\textsc{PromptImprovement}}
\label{alg:prompt}
\begin{algorithmic}[1]
\Require Prompts $P$, Existing rules $ER$, Handled failures $H$, LLM $M$, Domain contextual knowledge $C$, New prompts number $N$
\Ensure Optimized prompts $OP$, New induced rules $NR$, New handled failures $NH$
\State $OP\gets\varnothing$, $NR\gets\varnothing$
\Statex // Diversity-guided Prompt Generation
\State $SP \gets \textsc{SelectTop}(P, \textsc{Size}(P)-N)$ \Comment{Select the top $K$ prompts from $P$}
\State $D\gets$ generate $N$ different modification methods using $M$ 
\For{each $d$ in $D$}
\State $p\gets$ generate new prompt using $M$ based on $SP$ and $d$
\State $OP$.\textit{insert}($p$)
\EndFor
\State $OP \gets OP \cup SP$
\Statex // Failure-driven Rule Induction
\State $F \gets \textsc{ClusterFailureInfo}(SP)$ \Comment{Clustering by DBSCAN}
\State $F_i \gets \textsc{SampleRepresentativeCluster}(H, F)$ \Comment{Sample by Eq.~\ref{equ:weight}}
\State $(E, S) \gets \textsc{Reflection}(F_i)$ \Comment{Prompt $M$ to get explanations and solutions}
\State $R \gets \textsc{Summarize}(E, S)$ \Comment{Prompt $M$ to transform them into rules}
\For{each $r$ in $R$}
\If{$\textsc{Format(\textsc{SelectTop}(P, 1), $ER \cup r$, $C$)}> \textsc{Format(\textsc{SelectTop}(P, 1), $ER$, $C$)}$}
\State $NR$.\textit{insert}(r)
\EndIf
\EndFor
\State $NR \gets \textsc{SelectTop}(NR, 1) $, $NH \gets F_i$ \Comment{Select the best rule from $NR$ if $NR$ is not empty}
\State \textbf{return} $OP$, $NR$, $NH$
\end{algorithmic}
\end{algorithm*}

\subsubsection{Failure Information Selection} To identify shortcomings in current prompts, we propose to {delve into the failed test cases generated by current prompts and select their common errors.} 
Specifically, \tool first collects the failed test cases generated by the selected prompts $SP$ associated with the corresponding focal method and error messages. Then, \tool aggregates those failure information $F$ based on the typical DBSCAN~\cite{DBLP:conf/kdd/EsterKSX96} clustering algorithm (Lines 8). To determine which failures to address in each iteration, we employ a weighted sampling method. The weight of each cluster is based on two factors: its size and the similarity of its failure information to handled failures $H$ in previous iterations. A larger cluster size indicates a higher probability of {the failure type,}
so we assign a larger weight to it. 
As for the similarity with $H$, 
to prevent the model from getting stuck on the same difficult-to-solve issues, \tool measures the similarity between the failures in each cluster and those in $H$, and assigns a lower weight to clusters with higher similarity. Specifically, the weight is calculated as follows:

\begin{equation}
sim_i = 1-\max_{h \in H}(\frac{ED(f_i, h)}{\max(len(f_i), len(h))})
\end{equation}
\begin{equation}\label{equ:weight}
weight_i = \frac{size(f_i) \cdot sim_i}{\sum_{j=1}^{n} size(f_j) \cdot sim_j}
\end{equation}
where $ED( \cdot )$ denotes edit distance, {and} $size( \cdot )$ denotes the corresponding cluster size. $f_i$ and $T$ denote the failure information of the $i^{th}$ cluster's center sample and handled failures $H$ in previous iterations, respectively.

\subsubsection{Error Reflection} 
\begin{figure*}[t]
    \centering
    \begin{subfigure}[b]{0.31\textwidth}
        \centering
        \includegraphics[width=1\textwidth]{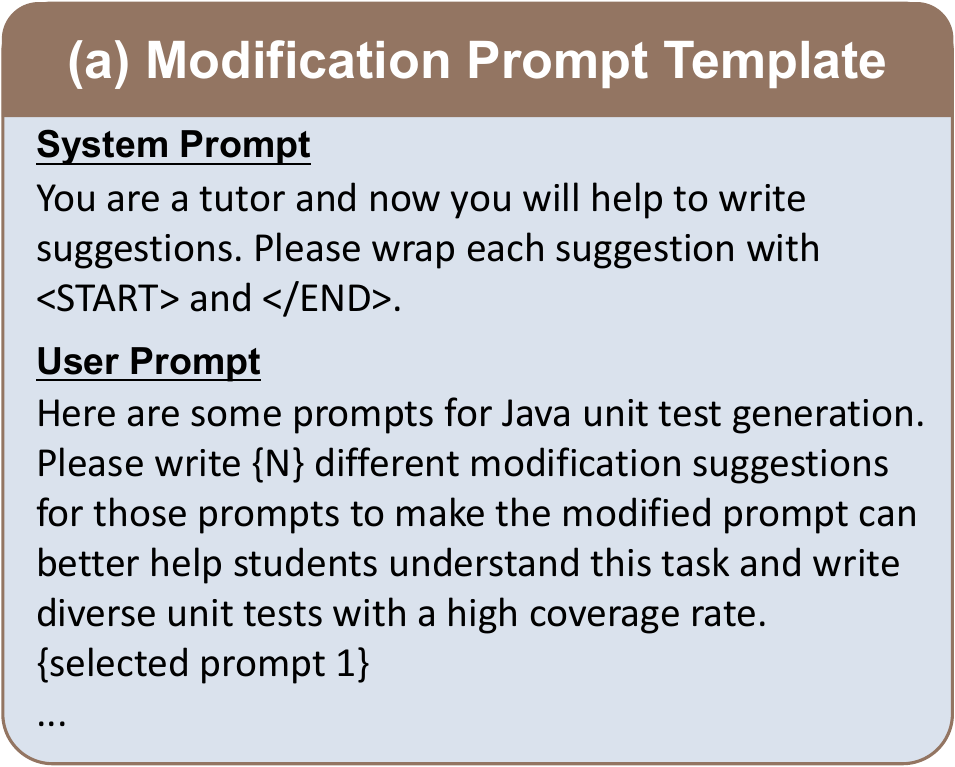}
      \end{subfigure}
      \hfill
      \begin{subfigure}[b]{0.31\textwidth}
        \centering
         \includegraphics[width=1\textwidth]{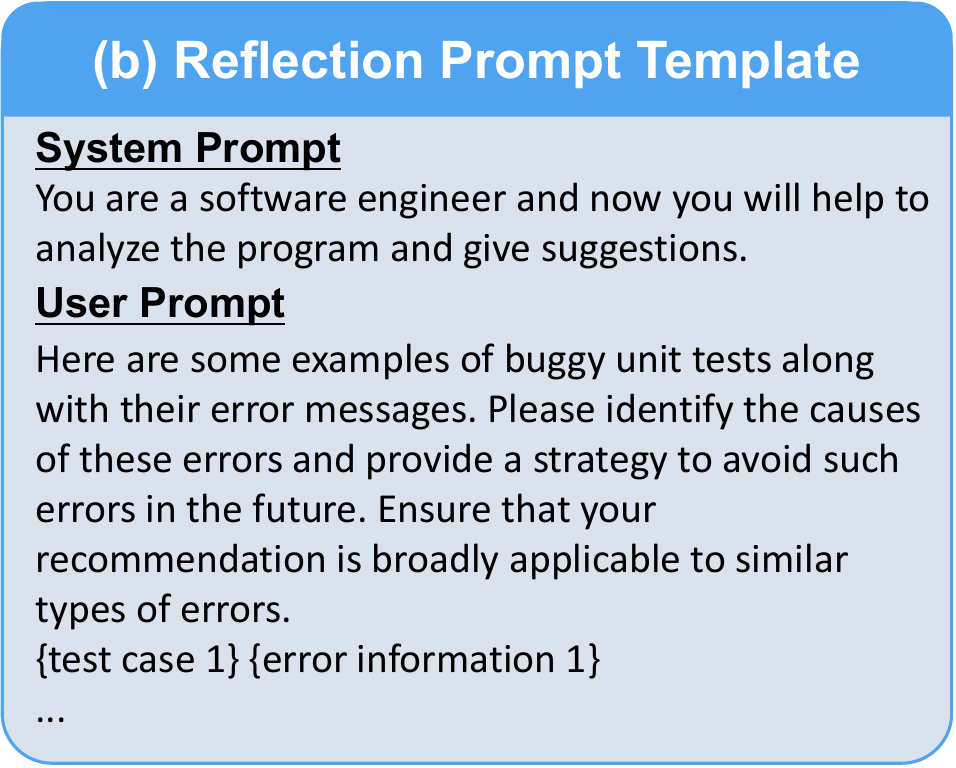}
      \end{subfigure}
      \hfill
      \begin{subfigure}[b]{0.31\textwidth}
        \centering
         \includegraphics[width=1\textwidth]{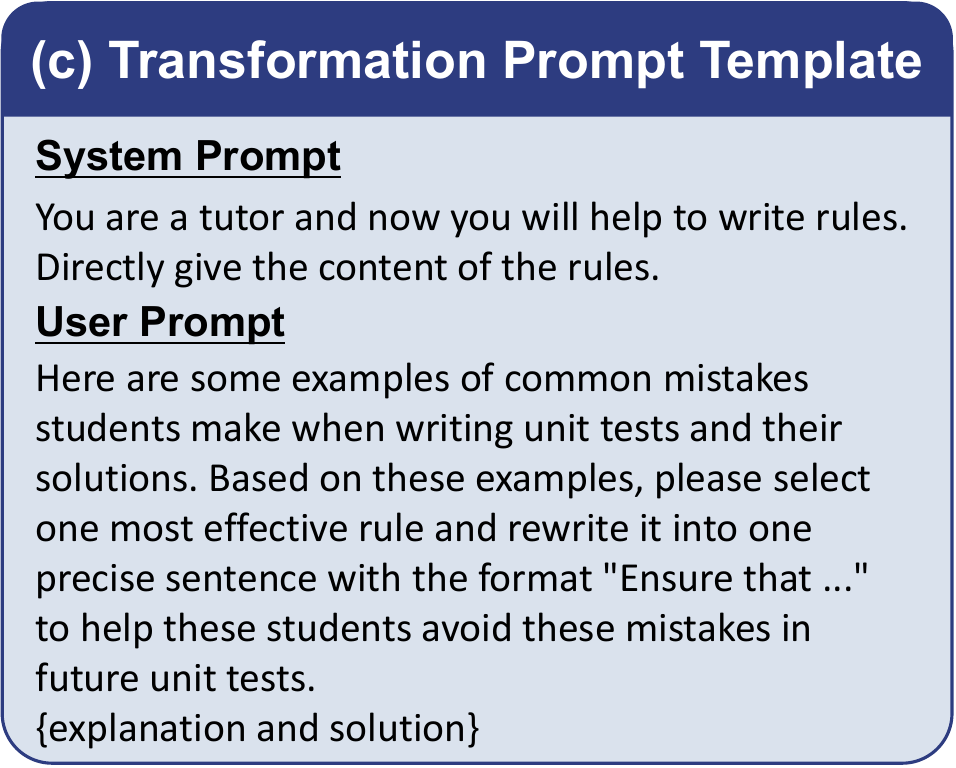}
      \end{subfigure}
    \caption{The prompt templates of \tool. The complete ones can be found in our replication package~\cite{replication}.}
    \label{fig:template}
\end{figure*}

With the selected failure information $F_i$, this part aims to enhance prompts by incorporating effective mitigation strategies
to prevent LLMs from repeating the same errors. First, 
{\tool chooses a few test cases whose failure information {exhibits}
the lowest Euclidean Distance to the cluster center of $F_i$}
{to construct the reflection prompt as depicted in Fig.~\ref{fig:template}} (b). The reflection prompt is then used to instruct the LLM $M$ to provide detailed explanations and solutions for the failure information (Lines 10). Additionally, to ensure that the solutions can be applied to more examples and not just the given ones, the reflection prompt also requires the model to remove example-specific information and make the solutions applicable to other similar cases. To avoid potential performance degradation brought by lengthy prompts~\cite{DBLP:journals/tacl/LiuLHPBPL24,needle}, we propose a soft incorporation of reflection outputs by converting them into {concise} rules. Specifically, \tool tasks 
{LLM $M$} 
with transforming the explanations and solutions into structured rules $R$ {based on {the}
transformation prompt template as shown in Fig.~\ref{fig:template}} (c) (Lines 11). 

\subsubsection{Rule Validation} 
To maintain the quality of the induced rules, this part aims at retaining only the most effective ones by validating each newly generated rule and incorporating the best one into the prompt. 
{To this end,} 
as shown in Lines 12 to 14 of Algorithm~\ref{alg:prompt}, \tool first constructs temporary prompts for each newly generated rule. The optimized {instruction} part 
of the temporary prompts is {from} the best-performing one in $P$, and the induced rules part of the temporary prompts includes both the existing rules $ER$ and each newly generated rule $r$. \tool then evaluates the performance of the temporary prompts on the sampled development set and {incorporates}
the 
{rule corresponding to the temporary prompt that achieves the highest performance into the final prompt (Lines 15). } 




\section{EXPERIMENTAL Setup}\label{sec:setup}

\subsection{Research Questions}
In the evaluation, we focus on the following four research questions:

\begin{enumerate}[label=\bfseries RQ\arabic*:,leftmargin=.5in]
    \item {How effective is \tool compared with existing prompt optimization methods?}
    \item {Is}
    \tool able to generate tailored prompts for different LLMs?
    \item What is the impact of each module on the performance of \tool?
    \item How does \tool's performance vary under different experimental settings?
\end{enumerate}

To study RQ1, we conduct a comprehensive evaluation of \tool by comparing it with four representative prompt optimization methods on three popular LLMs. For RQ2, we assess \tool's ability to generate LLM-tailored prompts for different LLMs by evaluating the performance of optimized prompts produced by \tool and manually designed prompts on different LLMs. For RQ3, we remove different modules of \tool to evaluate their individual contributions. For RQ4, we investigate \tool's performance under different experimental settings, including the number of seed prompts, the number of generated prompts per iteration $N$, and the maximum iteration number $I$.


\subsection{Datasets and Metrics}

\begin{table}
  \centering
    \caption{Statistics of the Defects4J benchmark.}
    \label{tab:dataset}
    \scalebox{1}{
    \begin{tabular}{llrrr}
    \toprule
    Project & Abbr. & Bug number & Focal class & Focal method \\
    \midrule
    Commons-Cli & Cli & 29 & 13 & 645 \\
    Commons-Csv & Csv & 15 & 5 & 373 \\
    Gson        & Gson & 17 & 15 & 220 \\
    Jfreechart  & Chart & 26 & 24 & 1,318 \\
    Commons-Lang& Lang & 60 & 28 & 2,712 \\
    \midrule
    All &  & 147 & 85 & 5,278 \\
    \bottomrule
    \end{tabular}
    }
\end{table}

We evaluate \tool on the widely-used Defects4J~\cite{DBLP:conf/issta/JustJE14} dataset. {Following previous studies~\cite{tufano2020unit,DBLP:journals/corr/abs-2302-10352}, we use five commonly used Java projects from this dataset including Apache Commons CLI, Apache Commons CSV, Google Gson, JFreeChart, and Apache Commons Lang. For each project, we use the fixed versions used by existing work~\cite{DBLP:journals/corr/abs-2302-10352} for evaluation.} These projects span diverse domains, including command-line interface, data processing, serialization, visualization, and utility libraries, respectively. {Table~\ref{tab:dataset} presents the overall information on the dataset and the detailed information such as specific versions and commit hashes can be found in our replication package~\cite{replication}.} As for evaluation metrics, we follow previous work~\cite{DBLP:conf/sigsoft/IvankovicPJF19,DBLP:journals/pacmse/Ryan0S00RR24} and adopt two most popular metrics to evaluate the performance of \tool and the baseline approaches:
\begin{itemize}
    \item \textbf{Line coverage (\%)} measures the percentage of code lines executed during testing. It checks whether each line of the source code is executed at least once, i.e., Line Coverage(\%) = $\frac{\text{Number of executed lines}}{\text{Total number of lines}} \times 100$. Only the lines covered by passed test cases are used for calculation.
    \item \textbf{Branch coverage (\%)} measures the percentage of branches executed during testing. It checks whether each branch in control structures is executed, i.e., Branch Coverage(\%) = $\frac{\text{Number of executed branches}}{\text{Total number of branches}} \times 100$. Only the branches covered by passed test cases are used for calculation.
\end{itemize}

\subsection{Baselines}\label{subsec:dataset}
To provide a comprehensive evaluation, we experiment on three popular LLMs and compare \tool with four representative prompt optimization methods, with details as below.

For LLMs, we select the following powerful LLMs in code-related tasks for evaluation:
\begin{itemize}
    \item \textbf{ChatGPT}~\cite{ChatGPT} is a popular model known for its versatile capabilities across various fields such as code generation. It is a closed-source model developed by OpenAI and we use the latest version gpt-3.5-turbo-0125 in our experiments.
    \item \textbf{Llama-3.1}~\cite{llama3} is a family of state-of-the-art open-source LLMs that have different sizes including 7B, 70B, and 405B. In this paper, we use the instruction-tuned Llama-3.1-70B-Instruct for experiments. 
    \item \textbf{Qwen2}~\cite{qwen2} is an open-source large language model that achieves promising in a variety of code intelligence tasks. It has a 128k context length to deal with project-level long code. Specifically, we choose Qwen2-72B-Instruct in this paper.
\end{itemize}

As for prompt optimization methods, we compare \tool with the basic prompt and four state-of-the-art prompt optimization methods:
\begin{itemize}
    \item \textbf{Basic} denotes the performance of the best seed prompt. It is used to measure how much improvements could prompt optimization methods to achieve. 
    \item \textbf{APE}~\cite{DBLP:conf/iclr/ZhouMHPPCB23} is a typical prompt optimization method that directly asks LLMs to generate variants for current prompts that can keep their semantic meanings in each iteration. 
    \item \textbf{OPRO}~\cite{DBLP:conf/iclr/Yang0LLLZC24} further incorporates the performance information and lets the LLM generate new prompts that can enhance the test accuracy based on existing prompts and their performance.
    \item \textbf{\textsc{EvoPrompt}}~\cite{DBLP:conf/iclr/Guo0GLS0L0Y24} is the state-of-the-art prompt optimization method that generates new prompts based on evolutionary operators. It has two versions: \textsc{EvoPrompt} (GA) and \textsc{EvoPrompt} (DE), which use the Genetic Algorithm, and Differential Evolution, respectively.
\end{itemize}

\subsection{Implementation}\label{sec:detail}
In our experiments, the number of seed prompts, the number of generated prompts per iteration $N$, and the maximum iteration number $I$ are set to 5, 2, and 5, respectively. The impact of different experimental settings is discussed in Section~\ref{subsec:rq4}. {We repeat \tool three times and report its average results and variance to eliminate the influence of sampling and fluctuations in LLM.}
During the prompt optimization stage, {we randomly sample ten bugs from the Defects4J benchmark as our development set $D_{dev}$ and use all bugs as test set $D_{test}$. We present the sampled development set $D_{dev}$ in our replication package~\cite{replication}.} To save manual efforts, we obtain the seed prompts automatically by ChatGPT and the existing Automatic Prompt Engineer method~\cite{DBLP:conf/iclr/ZhouMHPPCB23}. To ensure a fair comparison, we use the same development set and seed prompts for our tool and all baseline methods. The seed prompts, and all prompt templates used in our work can be found in our replication package~\cite{replication}. We conduct all experiments on an Ubuntu 20.04 server with a 112-core Intel Xeon Platinum CPU.
\section{Experimental Results}\label{sec:result}

\begin{table*}
    \centering
    \caption{Comparison with prompt optimization methods on ChatGPT. The number in ``()" denotes the standard deviation. 
    }
    \label{tab:rq1_gpt}
    \scalebox{1}{
    \begin{tabular}{c|ccccc|c}
    \toprule
    Projects & Chart & Cli & Csv & Gson & Lang & Average \\
    \midrule
    & \multicolumn{6}{c}{Line Coverage} \\
    \midrule
    Basic           & 47.41 & 36.76 & 37.49 & 22.42 & 57.58 & 45.56  \\
    APE            & 43.34 & 39.43 & 39.32 & 23.42 & 54.92 & 44.58 \\
    OPRO   & 44.52 & 42.49 & 38.96 & \textbf{25.71} & 53.70 & 45.13 \\
    \textsc{EvoPrompt} (GA) & 48.47 & 45.71 & 38.59 & 21.82 & 56.34 & 46.63 \\
    \textsc{EvoPrompt} (DE) & 49.60 & 42.88 & 39.87 & 24.17 & 53.39 & 45.89 \\
    \midrule
    \tool & \textbf{51.56} (0.66) & \textbf{58.88} (1.14) & \textbf{50.05} (1.02) & 25.17 (0.60) & \textbf{61.35} (0.42) & \textbf{53.80} (0.04)\\ 
    \midrule
    & \multicolumn{6}{c}{Branch Coverage} \\
    \midrule
    Basic           & 33.07 & 23.27 & 32.39 & 14.21 & 46.19 & 34.24  \\
    APE            & 33.72 & 24.30 & 34.32 & 15.53 & 44.59 & 34.25 \\
    OPRO   & 33.86 & 28.26 & 32.78 & \textbf{16.94} & 42.46 & 34.28 \\
    \textsc{EvoPrompt} (GA) & 34.36 & 31.64 & 34.18 & 14.53 & 45.07 & 35.88 \\
    \textsc{EvoPrompt} (DE) & 35.70 & 29.32 & 34.24 & 16.04 & 43.27 & 35.11 \\
    \midrule
    \tool & \textbf{38.68} (0.25) & \textbf{41.54} (1.58) & \textbf{39.53} (1.75) & 16.20 (0.40) & \textbf{51.11} (0.40) & \textbf{41.84} (0.26)\\ 
    \bottomrule
    \end{tabular}
    }
\end{table*}

\subsection{RQ1: Performance Evaluation}
To evaluate the effectiveness of \tool in test case generation, we compare it with four representative prompt optimization methods across three popular LLMs. Tables~\ref{tab:rq1_gpt}-\ref{tab:rq1_qwen} present the performance of \tool along with baseline methods on Defects4J. For each method, we provide the average performance across all bugs, as well as detailed average results for each project. Based on these results, we derive the following findings.

\textbf{Existing prompt optimization methods struggle to produce effective prompts for test case generation.}
By comparing the performance of the basic prompt and four baseline methods, we can observe that existing methods struggle to produce effective prompts for test case generation. Specifically, as shown in Table~\ref{tab:rq1_gpt}, the best-performing baseline, \textsc{EvoPrompt} (GA), can only achieve 1.07\% and 1.64\% improvements over the basic prompt in line coverage and branch coverage, respectively. Moreover, methods like APE and OPRO even perform worse than the basic prompt in terms of line coverage, with decreases of 0.98\% and 0.43\%, respectively. This suggests that simply combining and mutating existing prompts is difficult to produce effective prompts for test case generation.

\textbf{\tool achieves substantial improvement over baseline methods.}
As can be seen in Table~\ref{tab:rq1_gpt}-\ref{tab:rq1_qwen}, \tool considerably outperforms the baseline methods across all LLMs. For example, compared with the strongest baseline method, \textsc{EvoPrompt} (GA), \tool achieves an average improvement of 6.19\% and 5.03\% 
in line coverage and branch coverage, respectively. These results demonstrate the effectiveness of \tool in finding effective prompts within the vast search space. 

\textbf{The performance of different LLMs on different projects varies.}
By comparing the performance on different projects across different LLMs, we further observe that different LLMs tend to perform well on different projects. For instance, as shown in Table~\ref{tab:rq1_gpt}-\ref{tab:rq1_qwen}, although the overall performance of ChatGPT and Llama-3.1 with basic prompts are similar, their performance on individual projects exhibits large differences. Specifically, on the \textit{Lang} project, ChatGPT outperforms Llama-3.1 and Qwen2 by 4.69\% and 15.47\% in terms of line coverage, respectively; while on the \textit{Csv} project, the performance of ChatGPT is much worse than Llama-3.1 and Qwen2, with a decrease of 5.70\% and 3.24\% in terms of line coverage, respectively. {This indicates that different LLMs tend to excel in different {domains}
and also demonstrates the importance of building tailored prompts for different LLMs.}

\textbf{{\tool could achieve higher improvements on the projects that the seed prompts do not perform well.}}
At last, as shown in Table~\ref{tab:rq1_gpt}-\ref{tab:rq1_qwen}, we find that the improvements achieved by \tool on different projects also vary across different LLMs. For instance, in the \textit{Lang} project, the relative improvement on ChatGPT and Qwen2 are 6.55\% and 14.98\%, respectively; whereas in the \textit{Csv} project, the improvement on ChatGPT and Qwen2 are 33.50\% and 22.27\%, respectively. {These results demonstrate that \tool can achieve a higher increase on projects where LLMs do not excel, and it can provide directed improvements tailored to different LLMs.}

\begin{table*}
    \centering
    \caption{Comparison with prompt optimization methods on Llama-3.1. The number in ``()" denotes the standard deviation. 
    }
    \label{tab:rq1_llama}
    \scalebox{1}{
    \begin{tabular}{c|ccccc|c}
    \toprule
    Projects & Chart & Cli & Csv & Gson & Lang & Average \\
    \midrule
    & \multicolumn{6}{c}{Line Coverage} \\
    \midrule
    Basic & 46.52 & 45.99 & 43.19 & 22.81 & 52.89 & 45.93  \\
    APE & 45.05 & 46.26 & 42.17 & 21.56 & 50.98 & 44.70 \\
    OPRO & 45.45 & 44.39 & 42.81 & 23.39 & 54.09 & 45.95 \\
    \textsc{EvoPrompt} (GA) & 45.95 & 45.82 & 42.70 & 24.70 & 54.24 & 46.52 \\
    \textsc{EvoPrompt} (DE) & 45.70 & 44.96 & 43.24 & 22.77 & 52.29 & 45.34 \\
    \midrule
    \tool & \textbf{49.68} (0.21) & \textbf{51.83} (1.54) & \textbf{44.05} (0.19) & \textbf{26.38} (0.96) & \textbf{58.56} (1.49) & \textbf{50.59} (0.56)\\
    \midrule
    & \multicolumn{6}{c}{Branch Coverage} \\
    \midrule
    Basic & 35.46 & 28.55 & 36.92 & 16.83 & 42.73 & 35.06  \\
    APE & 34.54 & 28.49 & 35.52 & 16.37 & 41.58 & 34.22 \\
    OPRO & 34.85 & 26.61 & 37.25 & 16.86 & 43.20 & 34.80 \\
    \textsc{EvoPrompt} (GA) & 34.71 & 29.28 & 35.51 & 16.74 & 43.01 & 35.03 \\
    \textsc{EvoPrompt} (DE) & 34.82 & 29.28 & 36.95 & 17.00 & 42.62 & 35.07 \\
    \midrule
    \tool & \textbf{37.73} (0.68) & \textbf{35.06} (0.27) & \textbf{39.02} (1.14) & \textbf{19.36} (0.65) & \textbf{48.24} (1.69) &\textbf{ 39.50} (0.68)\\
    \bottomrule
    \end{tabular}
    }
\end{table*}

\begin{table*}
    \centering
    \caption{Comparison with prompt optimization methods on Qwen2. The number in ``()" denotes the standard deviation. 
    }
    \label{tab:rq1_qwen}
    \scalebox{1}{
    \begin{tabular}{c|ccccc|c}
    \toprule
    Projects & Chart & Cli & Csv & Gson & Lang & Average \\
    \midrule
    & \multicolumn{6}{c}{Line Coverage} \\
    \midrule
    Basic & 39.75 & 36.80 & 40.73 & 26.51 & 42.11 & 38.70 \\
    APE & 39.52 & 34.76 & 45.73 & 23.83 & 44.44 & 39.41 \\
    OPRO & 40.84 & 38.51 & 38.94 & 26.41 & 33.41 & 35.49 \\
    \textsc{EvoPrompt} (GA) & 39.08 & 36.49 & 37.50 & 21.23 & 43.56 & 38.17 \\
    \textsc{EvoPrompt} (DE) & 39.08 & 36.49 & 37.50 & 21.23 & 43.56 & 38.17 \\
    \midrule
    \tool & \textbf{44.37} (3.56) & \textbf{47.56} (0.40) &\textbf{ 49.80} (3.48) & \textbf{29.75} (1.29) & \textbf{48.42} (2.16)  & \textbf{45.51} (1.28)\\
    \midrule
    & \multicolumn{6}{c}{Branch Coverage} \\
    \midrule
    Basic & 30.88 & 23.79 & 32.55 & 18.18 & 30.55 & 28.05 \\
    APE & 31.07 & 21.35 & 36.22 & 16.79 & 33.27 & 28.92 \\
    OPRO & \textbf{32.47} & 26.76 & 29.58 & \textbf{19.92} & 25.26 & 26.66 \\
    \textsc{EvoPrompt} (GA) & 30.81 & 22.16 & 29.71 & 13.84 & 33.14 & 27.98 \\
    \textsc{EvoPrompt} (DE) & 30.81 & 22.16 & 29.71 & 13.84 & 33.14 & 27.98 \\
    \midrule
    \tool & 30.58 (3.26) & \textbf{31.73} (1.44) & \textbf{36.30} (2.83) & 18.98 (2.01) & \textbf{37.11} (1.51)  & \textbf{32.71} (1.43)\\
    \bottomrule
    \end{tabular}
    }
\end{table*}

\begin{table*}[t]
    \centering
    \caption{Evaluation of \tool in generating  tailored prompts for different LLMs.}
    \label{tab:cross}
    \scalebox{1}{
    \begin{tabular}{c|ccc|ccc}
    \toprule
     Approach & ChatGPT & Llama-3.1 & Qwen2 & ChatGPT & Llama-3.1 & Qwen2  \\
    \midrule
    & \multicolumn{3}{c|}{Line Coverage} & \multicolumn{3}{c}{Branch Coverage} \\
    \midrule
    ChatGPT's final prompt & \textbf{53.80} & 41.92 & 35.98 & \textbf{41.84} & 32.31 & 26.87  \\
    Llama-3.1's final prompt & 51.35 & \textbf{50.59} & 44.94 & 40.05 & \textbf{39.50} & \textbf{34.43} \\
    Qwen2's final prompt & 51.14 & 43.98 & \textbf{45.51}& 38.39 & 32.97 & 32.71 \\ 
    Manually-designed prompt & 48.55 & 48.46 & 42.85  & 37.55 & 37.60 & 31.88 \\ 
    \bottomrule
    \end{tabular}
    }
\end{table*}

\begin{tcolorbox}[width=\linewidth,boxrule=0pt,top=1pt, bottom=1pt, left=1pt,right=1pt, colback=gray!20,colframe=gray!20]
\textbf{Answer to RQ1:} 
\tool effectively enhances prompts for test case generation. It consistently outperforms all baseline methods across various LLMs, achieving a 6.19\% higher line coverage rate and a 5.03\% higher branch coverage rate compared to the strongest baseline.
\end{tcolorbox}

\subsection{RQ2: LLM-Tailored Prompt Generation Evaluation}
In this RQ, we study whether \tool could generate tailored prompts for different LLMs. To achieve this, we evaluate the performance of the three final prompts obtained by three models on each model. Additionally, we also compare the prompt used in~\cite{DBLP:journals/pacmse/Yuan0DW00L24} to validate whether the prompt built by \tool could outperform the manually-designed prompt. The experimental results are depicted in Table~\ref{tab:cross}. The detailed results on each project can be found in our replication package~\cite{replication}. Based on the results, we have the following observations:

\textbf{The performance of different prompts varies a lot.} 
By comparing the performance of each final prompt on different models. We can find that the performance of different prompts on the same LLM varies a lot. Specifically, the line coverage rate and branch coverage rate of Llama-3.1 on different final prompts range from 41.92\%-50.59\% and 32.31\%-39.50\%, respectively, which further demonstrates the importance of automated generating tailored prompts for different LLMs.

\textbf{\tool effectively produces tailored prompts for different LLMs.} 
As in Table~\ref{tab:cross}, we can observe that each model tends to achieve the best performance on their own final prompt. For example, the performance of ChatGPT's final prompt outperforms the final prompt obtained by Llama-3.1 and Qwen2 by 2.45\% and 2.66\% in terms of line coverage on ChatGPT. This indicates that \tool could effectively produce tailored prompts for each LLM.

{\textbf{Prompts optimized by \tool outperform manually-designed prompts.} Additionally, the prompt obtained by \tool also outperforms the line coverage of manually-designed prompt by 5.25\%, 2.13\%, and 2.66\% on ChatGPT, Llama-3.1, and Qwen2, respectively. These results demonstrate \tool's efficacy in automatically crafting effective, LLM-tailored prompts.}


\begin{tcolorbox}[width=\linewidth,boxrule=0pt,top=1pt, bottom=1pt, left=1pt,right=1pt, colback=gray!20,colframe=gray!20]
\textbf{Answer to RQ2:} 
The performance of different prompts varies a lot and \tool could effectively produce tailored prompts for different LLMs.
\end{tcolorbox}

\begin{table*}[t]
    \centering
    \caption{Ablation Study of \tool.}
    \label{tab:ablation}
    \scalebox{1}{
    \begin{tabular}{cc|cc}
    \toprule
    \multicolumn{2}{c|}{Approach} & Line Coverage & Branch Coverage \\
    \midrule
    \multirow{5}{*}{ChatGPT} & \multicolumn{1}{|c|}{\tool}  & \textbf{53.80}  & \textbf{41.84} \\
    & \multicolumn{1}{|c|}{w/o Domain contextual knowledge extraction} & 44.16  & 33.31 \\
    & \multicolumn{1}{|c|}{w/o Diversity-guided prompt generation} & 45.59  & 34.04 \\
    & \multicolumn{1}{|c|}{w/o Failure-driven rule induction} & 46.86  & 37.08 \\
    & \multicolumn{1}{|c|}{Only domain contextual knowledge extraction} & 48.03 & 35.76 \\
    \midrule
    \multirow{5}{*}{Llama-3.1} & \multicolumn{1}{|c|}{\tool}  & \textbf{50.59}  & \textbf{39.50} \\
    & \multicolumn{1}{|c|}{w/o Domain contextual knowledge extraction} & 45.37  & 34.73 \\
    & \multicolumn{1}{|c|}{w/o Diversity-guided prompt generation} & 49.68  & 38.35 \\
    & \multicolumn{1}{|c|}{w/o Failure-driven rule induction} & 49.73  & 38.88 \\
    & \multicolumn{1}{|c|}{Only domain contextual knowledge extraction} & 49.48 & 37.95 \\
    \midrule
    \multirow{5}{*}{Qwen2} & \multicolumn{1}{|c|}{\tool}  & \textbf{45.51}  & \textbf{32.71} \\
    & \multicolumn{1}{|c|}{w/o Domain contextual knowledge extraction} & 42.12  & 29.73 \\
    & \multicolumn{1}{|c|}{w/o Diversity-guided prompt generation} & 43.81  & 32.10 \\
    & \multicolumn{1}{|c|}{w/o Failure-driven rule induction} & 43.92  & 31.58 \\
    & \multicolumn{1}{|c|}{Only domain contextual knowledge extraction} & 40.56 & 30.21 \\
    \bottomrule
    \end{tabular}
    }
\end{table*}

\subsection{RQ3: Ablation Study}
We conduct ablation studies to validate the effectiveness of each module in our method, i.e. domain contextual knowledge extraction, diversity-guided prompt generation, and failure-driven rule induction. The average results for each method are presented in Table~\ref{tab:ablation}, with detailed results for each project available in our replication package~\cite{replication}.

\textbf{Domain contextual knowledge extraction.} We conduct this experiment by removing the cross-file context information in the final prompt. As can be seen in Table~\ref{tab:ablation}, excluding the cross-file context information dramatically degrades performance across all LLMs. Specifically, the branch coverage rate drops by 8.53\%, 4.77\%, and 2.98\% on ChatGPT, Llama-3.1, and Qwen2, respectively.  These results demonstrate the effectiveness of integrating project context information to help LLMs generate accurate test cases.

\textbf{Diversity-guided prompt generation.} To validate the effectiveness of diversity-guided prompt generation, we experiment by replacing the optimized instruction part of the final prompt with the one produced by the best baseline method. As shown in Table~\ref{tab:ablation}, removing the diversity-guided prompt generation leads to a consistent drop in all tasks and metrics. For example, the line coverage rate decreases by 8.21\%, 1.01\%, and 1.70\% on ChatGPT, Llama-3.1, and Qwen2, respectively, which demonstrates the importance of prompt diversity in the search space exploration process.

\textbf{Failure-driven rule induction.} We conduct this experiment by removing the induced rules in the final prompt. From Table~\ref{tab:ablation}, we can observe that without failure-driven rule induction, the performance of \tool drops a lot across all LLMs. Specifically, in ChatGPT, the line coverage decreases by 6.94\% and branch coverage by 4.76\%, respectively. This indicates the benefits of using LLM-induced rules to guide the optimization process and avoid LLMs making recurring errors. We further show some cases in Section~\ref{sec:case} for illustration.

\textbf{Only domain contextual knowledge extraction.} {As the domain contextual knowledge extraction module provides the most significant performance gains, we further evaluate how much could this module only bring to the basic prompt to ensure fairness in comparison. We conduct this experiment by removing both the diversity-guided prompt generation module and the failure-driven rule induction module. As shown in Table~\ref{tab:ablation}, removing both these two parts lead to substantial performance to \tool. Specifically, solely involving the domain contextual knowledge extraction can only bring limited improvement over the basic prompt, i.e., improving the line coverage and branch coverage for Qwen2 by 1.86\% and 2.16\%, respectively. Its performance still falls behind \tool by a large margin, which indicates that simply combining basic prompt and the context information without further optimization can not achieve satisfactory performance.}

\begin{tcolorbox}[width=\linewidth,boxrule=0pt,top=1pt, bottom=1pt, left=1pt,right=1pt, colback=gray!20,colframe=gray!20]
\textbf{Answer to RQ3:} 
All modules in \tool contribute to the performance. {Removing the domain contextual knowledge extraction part leads to the largest performance decreases.}
\end{tcolorbox}

\begin{figure*}[t]
    \centering
    \begin{subfigure}[b]{0.33\textwidth}
        \centering
        \includegraphics[width=1\textwidth]{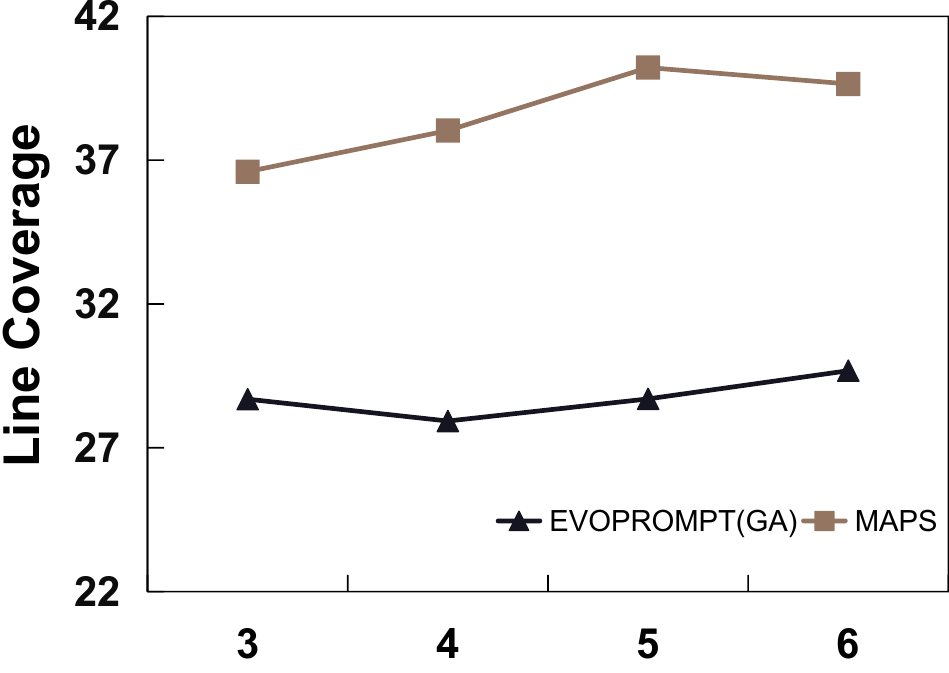}
        \caption{Number of seed prompts.}
      \end{subfigure}
      \hspace{1cm}
      \begin{subfigure}[b]{0.33\textwidth}
        \centering
         \includegraphics[width=1\textwidth]{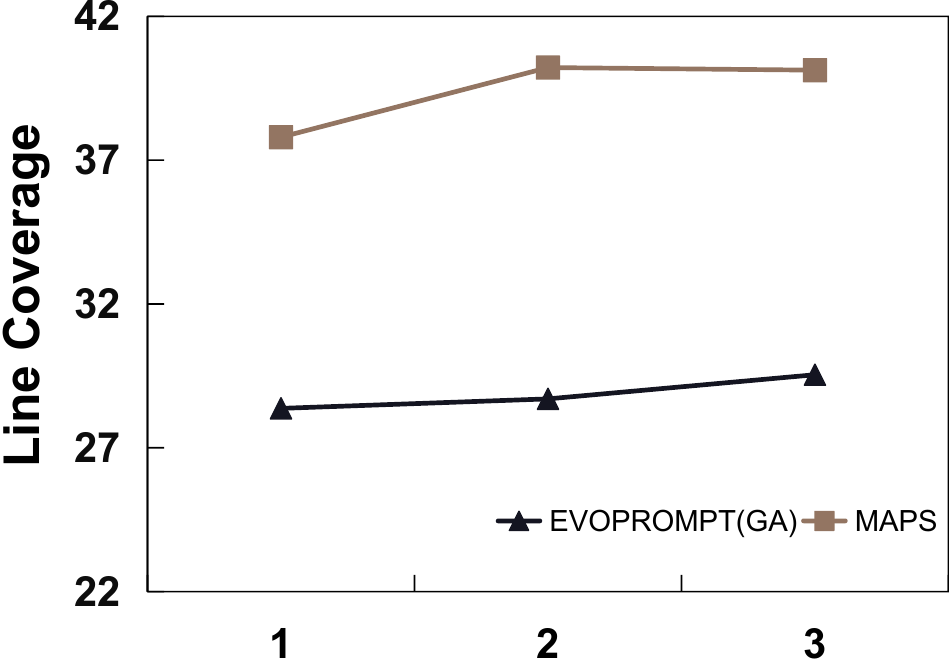}
        \caption{Number of generated prompts.}
      \end{subfigure}
    \caption{Parameter analysis of number of seed prompts and generated prompts on ChatGPT.}
    \label{fig:para}
\end{figure*}

\begin{figure*}[t]
    \centering
    \begin{subfigure}[b]{0.32\textwidth}
        \centering
        \includegraphics[width=1\textwidth]{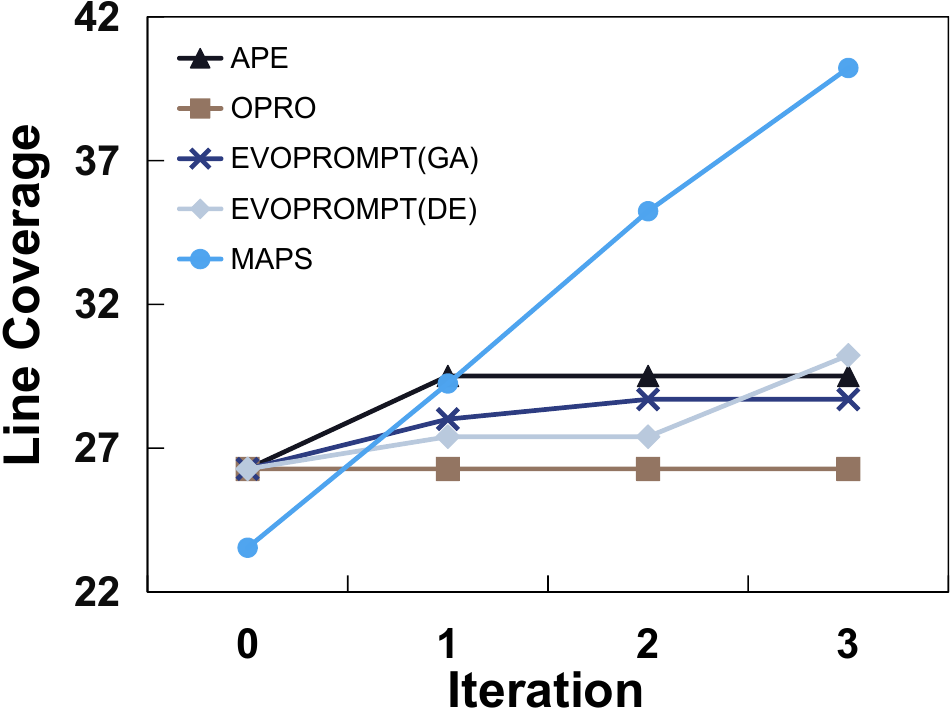}
        \caption{ChatGPT.}
      \end{subfigure}
      \hfill
      \begin{subfigure}[b]{0.32\textwidth}
        \centering
         \includegraphics[width=1\textwidth]{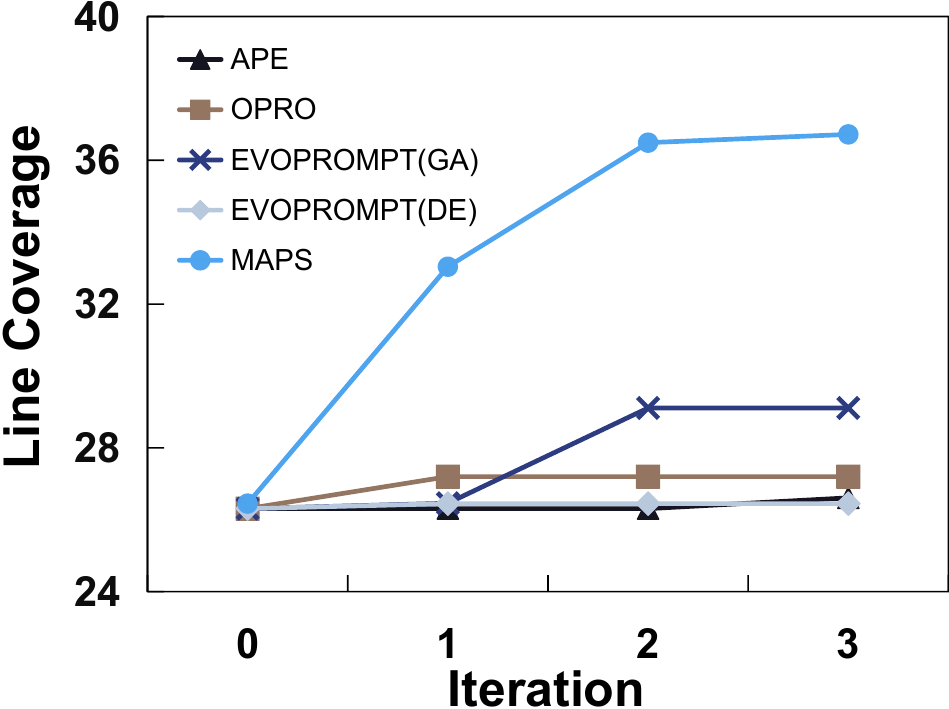}
        \caption{Llama-3.1.}
      \end{subfigure}
      \hfill
      \begin{subfigure}[b]{0.32\textwidth}
        \centering
         \includegraphics[width=1\textwidth]{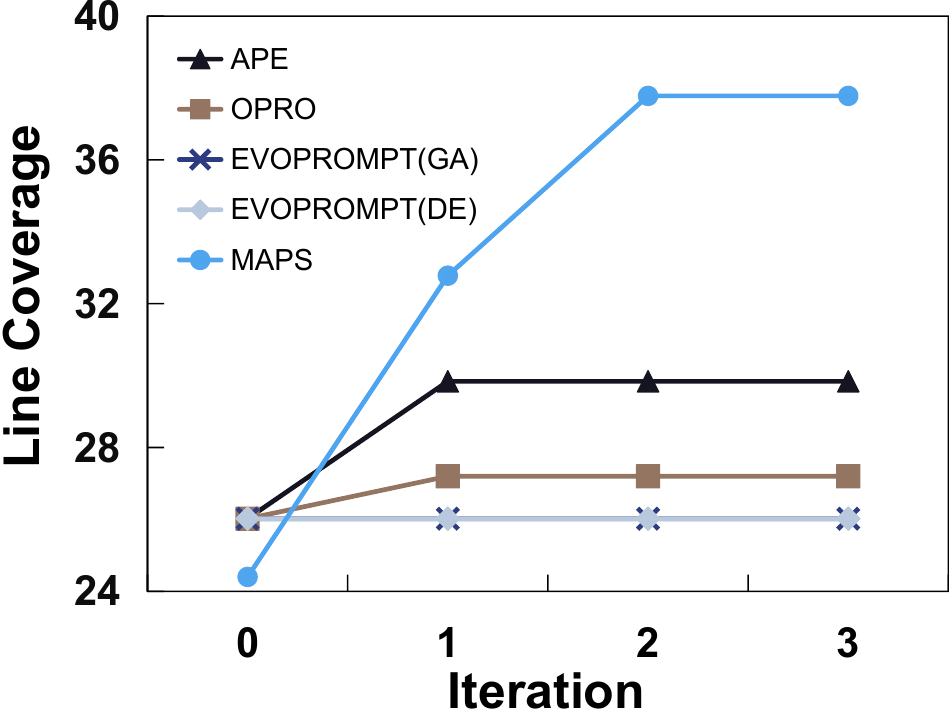}
        \caption{Qwen2.}
      \end{subfigure}
    \caption{Parameter analysis of the Iteration number.}
    \label{fig:para1}
\end{figure*}

\subsection{RQ4: Parameter Analysis}~\label{subsec:rq4} 
In this section, we study how different experimental settings affect the performance of \tool and baseline methods, including the number of seed prompts, the number of generated prompts $N$, and the maximum iteration number $I$. As these parameters primarily influence the prompt optimization process, we report their performance on the development set in this section. In each study, we vary only the parameter under analysis and keep others constant. {For the analysis on number of seed prompts and generated prompts, we only present the results on ChatGPT};
the complete results are available in our replicate package~\cite{replication}. 

\textbf{Number of seed prompt.} We conduct experiments to evaluate how \tool and baseline methods perform under different numbers of seed prompts. Specifically, we use the best-performing baseline \textsc{EvoPrompt} (GA) and set the number of seed prompts to 3, 4, 5, and 6 respectively. From Fig.~\ref{fig:para} (a), we can observe that \tool consistently achieves better performance across different numbers of seed prompts. Additionally, by comparing the performance under different numbers of seed prompts, we can find that \tool and \textsc{EvoPrompt} (GA) tend to achieve better performance with a larger number of seed prompts, and the improvements over five seed prompts are marginal. Therefore, we set the number of seed prompts to five in this paper.

\textbf{Number of generated prompt.} We also study the effect of a number of generated prompts by varying it from 1 to 3. As shown in Fig.~\ref{fig:para} (b), \tool consistently achieves better performance across different numbers of generated prompts. While a larger number of generated prompts can lead to better performance, it also increases costs. Therefore, we set the number of generated prompts to two in this experiment.

\textbf{Maximum iteration number.} In this study, we vary the number of maximum iteration number from 1 to 3 and investigate the performance of prompts in each iteration. We present the results of the best prompt generated by \tool and each baseline method on the development set. Iteration 0 represents the performance of the basic prompt without optimization by \tool. As shown in Fig.~\ref{fig:para1}, \tool achieves the best performance in most cases. Specifically, \tool outperforms baseline methods by at least 7.94\% in line coverage when the maximum iteration number is set to three. {Additionally, due to low prompt diversity, baseline methods tend to converge to local optima in the first iteration and fail to achieve further improvement. In contrast, \tool continually enhances performance with each iteration.}

\begin{tcolorbox}[width=\linewidth,boxrule=0pt,top=1pt, bottom=1pt, left=1pt,right=1pt, colback=gray!20,colframe=gray!20]
\textbf{Answer to RQ4:} 
\tool consistently achieves the best performance across different parameter settings. {Our hyperparameter settings, with the number of seed prompts set to 5, $N$ to 2, and $I$ to 5, achieve effective results}.
\end{tcolorbox}

\section{Discussion}\label{sec:discuss}

\begin{figure*}[t]
    \centering
    \begin{subfigure}[b]{0.39\textwidth}
        \centering
        \includegraphics[width=1\textwidth]{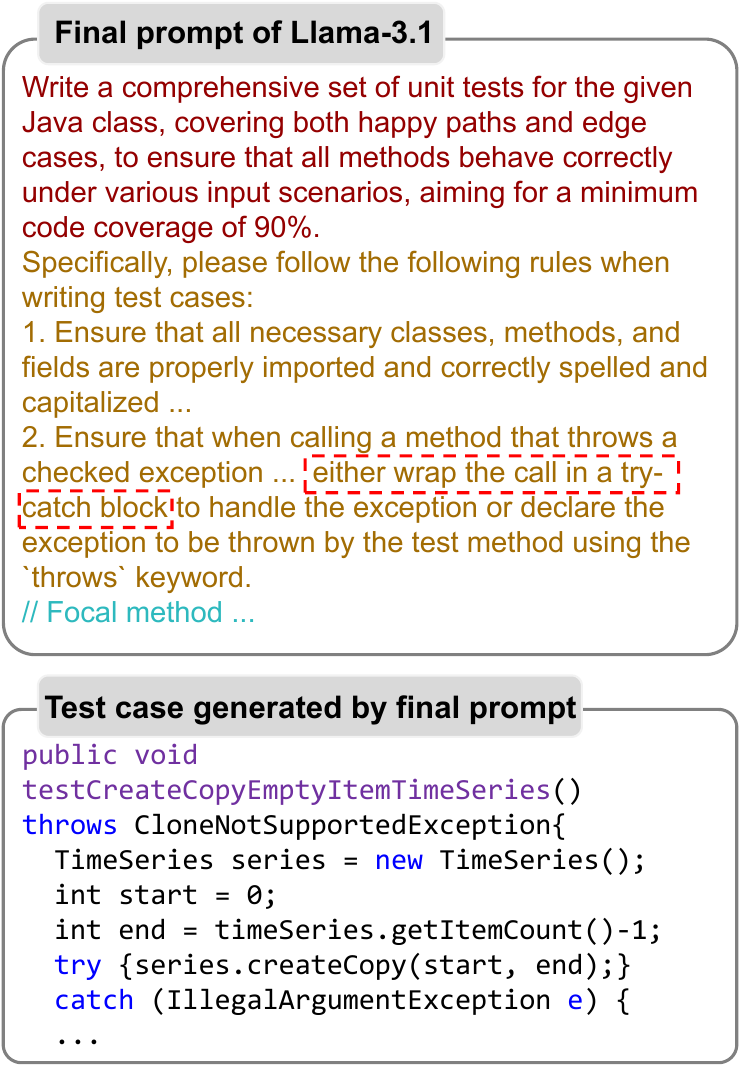}
        \caption{Case study on Llama-3.1.}
      \end{subfigure}
      \hspace{0.5cm}
      \begin{subfigure}[b]{0.495\textwidth}
        \centering
         \includegraphics[width=1\textwidth]{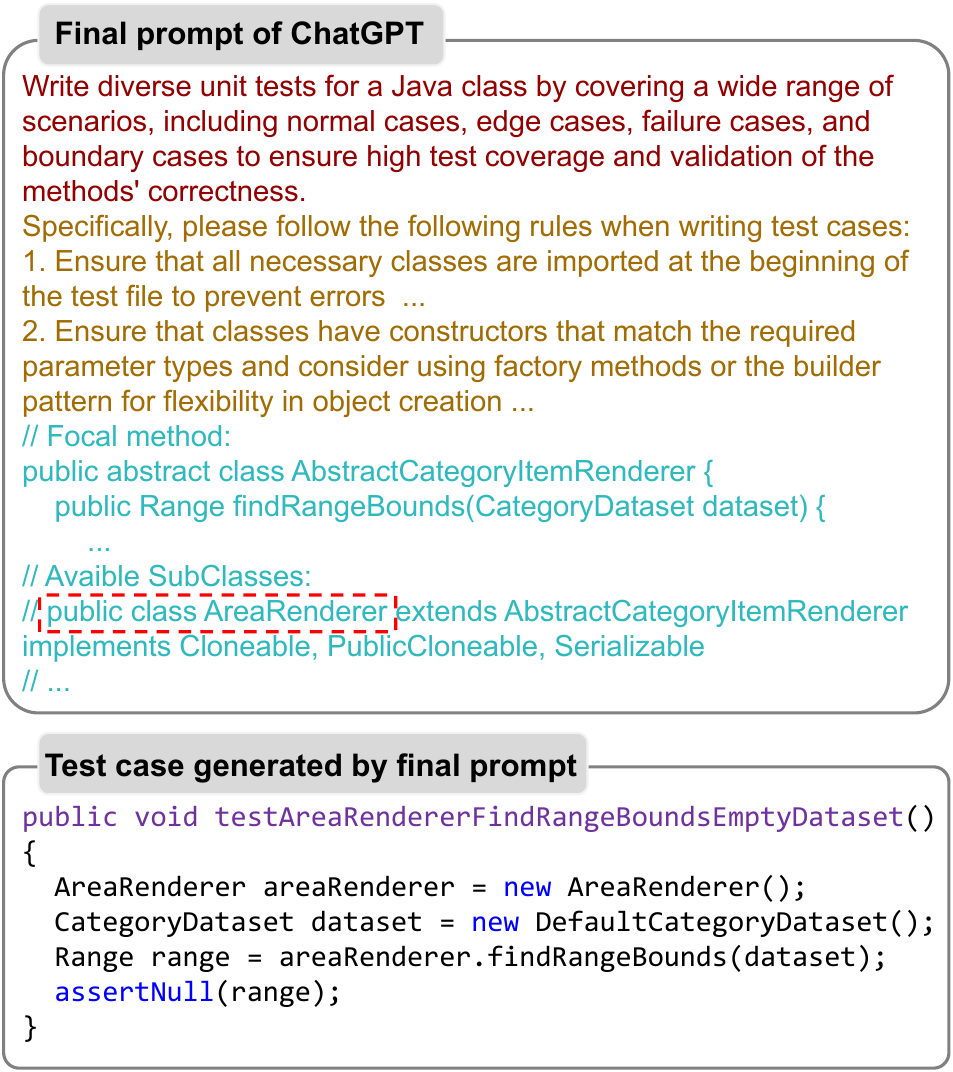}
        \caption{Case study on ChatGPT.}
      \end{subfigure}
    \caption{Two cases showing the difference of optimized {prompts}
    from different models and how the optimized prompt help generate correct test case.}
    \label{fig:case}
\end{figure*}
\subsection{Case study}\label{sec:case}
To better understand how \tool improves test case generation, we present two examples of the final prompts created by \tool and the resulting test cases from these final prompts. First, Fig~\ref{fig:case} (a) shows the final prompt for Llama-3.1, along with the generated test case based on the focal method in Listing~\ref{lst:example2}. We can find that by following the second induced rule, Llama-3.1 correctly generates a test case that uses the ``\texttt{try \{...\} catch (IllegalArgumentException e)}'' in the test method. Second, Fig~\ref{fig:case} (b) illustrates another example using ChatGPT, where the focal method is taken from Listing~\ref{lst:example3}. Compared to the incorrect test case generated by the baseline prompt in Listing~\ref{lst:example3}, the cross-file contextual knowledge in the optimized prompt allows the model to correctly initialize the ``\texttt{AbstractCategoryItemRenderer}'' class. Additionally, by comparing two final prompts obtained by \tool, we observe that the induced rules for Llama-3.1 and ChatGPT are different. The first {rules}
of these two methods are similar, but Llama-3.1's second rule focuses on exception handling, whereas ChatGPT's concerns method parameters. This indicates that these models tend to make different types of errors, and \tool can effectively introduce tailored rules for different LLMs.

Moreover, by calculating the average edit distance of prompts obtained in each optimization iteration by \tool, we find that it generates more diverse prompts during the optimization process. Specifically, the average edit distance of prompts from \tool is 27.0, remarkably larger than that of OPRO, which averages only 9.3 edits, as shown in Table~\ref{tab:example1}. This further demonstrates \tool's effectiveness in generating diverse prompts during the optimization process.

\subsection{Comparison with Other Methods}
To comprehensively study the advantages and limitations of LLMs-based test case generation methods compared with traditional approaches and previous deep learning-based approaches, we compare ChatGPT+\tool with two baseline methods including Randoop~\cite{DBLP:conf/oopsla/PachecoE07} and A3Test~\cite{DBLP:journals/corr/abs-2302-10352}. Randoop~\cite{DBLP:conf/oopsla/PachecoE07} is a widely used automated software testing tool that employs random fuzzing on unit APIs to construct prefixes that lead the unit into noteworthy states. A3Test~\cite{DBLP:journals/corr/abs-2302-10352} is a state-of-the-art non-LLM-based deep learning model that fine-tunes PLBART~\cite{DBLP:conf/naacl/AhmadCRC21} for test case generation. For Randoop, we reproduce it based on the ``\texttt{gen\_tests.pl}'' script provided in Defects4J. For A3Test, we reproduce the results based on A3Test's replicate package~\cite{a3testreplication}.


\begin{table}[t]
  \centering
    \caption{Comparison of Randoop, A3Test, ChatTESTER, and \tool.}
    \label{tab:other}
    \scalebox{1}{
    \begin{tabular}{c|cc}
    \toprule
    Projects & Line Coverage & Branch Coverage \\
    \midrule
    Randoop & 49.51 & 34.45 \\
    A3Test & 34.11 & 15.72 \\
    \midrule
    ChatGPT+Basic & 45.56 & 34.24 \\
    ChatGPT+\tool & \textbf{53.80} & \textbf{41.84} \\
    \bottomrule
    \end{tabular}
    }
\end{table}

Table~\ref{tab:other} presents the experimental results in terms of line coverage and branch coverage. Compared to three baseline methods, ChatGPT+\tool achieves the highest line coverage (i.e., 53.80\%) and branch coverage (i.e., 41.84\%), outperforming traditional methods by at least 4.29\% and 7.39\%, respectively. This demonstrates \tool's effectiveness in helping LLMs generate test cases with high coverage. Besides, when comparing the performance of Randoop, ChatGPT+basic, and ChatGPT+\tool, we can find that the performance of ChatGPT+basic is lower than traditional method Randoop, meaning that simply prompting LLMs can not achieve satisfactory results, while \tool can produce suitable prompts for LLMs and make LLMs outperform traditional methods. 

Although these are also other LLM-based methods such as ChatUniTest~\cite{DBLP:journals/pacmse/Yuan0DW00L24}, we do not compare with them because our research is orthogonal to them. \tool focuses on optimizing tailored prompts for test case generation. It can be further combined with existing methods such as incorporating static information~\cite{DBLP:journals/pacmse/Ryan0S00RR24} or multi-turn refinement method~\cite{DBLP:journals/pacmse/Yuan0DW00L24} and achieve better performance.

\subsection{Threats to Validity}
We identify two main threats to the validity of our study:

\textbf{Limited LLMs}. Given the rapid development of large language models, some models are not covered in this paper. To mitigate this issue, we select the three most representative and popular LLMs that contain both open-source models and closed-source models. Additionally, \tool is model-agnostic and does not require access to the model's parameters. Therefore, we believe \tool can also achieve improvements on other LLMs.

\textbf{Limited Programming Languages}. In this paper, we conduct experiments using the Defects4J benchmark, which only contains Java projects. This benchmark is popular and widely used in previous work. Furthermore, our method is language-agnostic and can be easily adapted to other programming languages. In the future, we plan to conduct experiments on more datasets including those with languages such as Python.

\section{Related work}\label{sec:related}

\subsection{Automatic Prompt Optimization}
Automatically discovering optimal prompts has emerged as an important challenge in the era of LLMs~\cite{DBLP:conf/emnlp/PryzantI0L0023,DBLP:conf/iclr/ZhouMHPPCB23}. Most existing methods follow an iterative prompt optimization process. They start with a set of seed prompts and iteratively synthesize new prompt candidates, evaluating their performance to select the top ones for the next iteration. 
For example, APE~\cite{DBLP:conf/iclr/ZhouMHPPCB23} is a typical prompt optimization method that directly asks LLMs to generate variants of current prompts while maintaining their semantic meanings in each iteration. OPRO~\cite{DBLP:conf/iclr/Yang0LLLZC24} further incorporates the performance information and lets the LLM generate new prompts that can enhance the test accuracy based on existing prompts and their performance. \textsc{EvoPrompt}~\cite{DBLP:conf/iclr/Guo0GLS0L0Y24} is the state-of-the-art prompt optimization method that generates new prompts based on evolutionary operators. It has two versions: \textsc{EvoPrompt} (GA) and \textsc{EvoPrompt} (DE), which use the Genetic Algorithm, and Differential Evolution, respectively. 
{Different from those works, this paper focuses on LLM-tailored prompt optimization for test case generation and investigates improving the exploration of the search process. 
}

\subsection{Test Case Generation}
Traditional methods like Randoop~\cite{DBLP:conf/oopsla/PachecoE07} utilize random fuzzing on unit APIs to construct prefixes that lead the unit into noteworthy states. Evosuite~\cite{DBLP:conf/sigsoft/FraserA11} is a search-based test generation strategy that employs evolutionary algorithms to autonomously craft test suites for Java classes aimed at improving coverage rate. A series of recent studies~\cite{tufano2020unit,DBLP:conf/icse/YuLSR00L0W22,DBLP:conf/kbse/SunLYLZ23} have employed deep learning techniques by training models to convert target methods into their corresponding test cases or assertions. A series of recent studies~\cite{tufano2020unit,DBLP:conf/icse/GaoMG000L24} have employed deep learning techniques for test case generation by formulating the test case generation as a neural machine translation task and train models to convert target methods into their corresponding test cases or assertions. 
For example, AthenaTest~\cite{tufano2020unit} fine-tunes BART~\cite{DBLP:conf/acl/LewisLGGMLSZ20} on a dataset designed for test generation. A3Test~\cite{DBLP:journals/corr/abs-2302-10352} further incorporates assertion knowledge and a test signature verification mechanism for achieving better results. 
Recently, leveraging advancements in LLMs, test case generation approaches based on LLMs have also been proposed and shown promising results. For example, CodaMOSA~\cite{DBLP:conf/icse/LemieuxILS23} leverages LLMs to provide example test cases for under-covered functions when search-based testing hits a coverage stall. ChatTESTER~\cite{DBLP:journals/pacmse/Yuan0DW00L24} incorporates ChatGPT along with an iterative test refiner to generate tests. {Different from those works, our method serves as the first LLM-tailored prompt generation method for test case generation and can be further combined with existing methods to enhance their performance. Besides, our method aims at to directly avoid generating low-quality test cases with an optimized prompt instead of time-consuming  multi-iteration generation and fixing during test time.}

\subsection{LLMs for Software Engineering}
Large Language Models have recently been widely adopted for various software engineering tasks due to their impressive performance in both code generation and understanding~\cite{DBLP:conf/icse/GaoZGW23,DBLP:journals/pacmse/Yang0YK0LHMJ024,DBLP:journals/tosem/GaoGHZNXL23,DBLP:conf/ccs/LiWWG23}. For example, Yuan et al.~\cite{DBLP:journals/pacmse/Yuan0DW00L24} evaluate the performance of ChatGPT for test case generation and improve it by iterative test refiner. Gao et al.~\cite{DBLP:conf/kbse/GaoWGWZL23} investigate how to set the in-context demonstration for ChatGPT for code summarization and code generation tasks. CHATRepair~\cite{DBLP:journals/corr/abs-2304-00385} iteratively evaluates programs on test cases and feeds the error messages to LLMs for further patch generation. Self-edit~\cite{DBLP:conf/acl/ZhangLLLJ23} utilizes compiler error messages to enhance the correctness of code generation. Li et al.~\cite{DBLP:conf/icse/LiW0L0WG024} investigates the feasibility of slicing commercial black-box LLMs using medium-sized backbone models. SBLLM~\cite{gao2024search} combines search-based methods and LLMs to iteratively improve code efficiency.
DeepSeek-Coder~\cite{DBLP:journals/corr/abs-2401-14196} is an open-source Mixture-of-Experts (MoE) code language model that achieves state-of-the-art performance across various code intelligence tasks. 
StarCoder 2~\cite{DBLP:journals/corr/abs-2402-19173} is an advanced LLM trained in 600+ programming languages. It is trained on the Stack2~\cite{DBLP:journals/corr/abs-2402-19173} dataset and natural language text from Wikipedia, Arxiv, and GitHub issues. Magicoder~\cite{DBLP:journals/corr/abs-2312-02120} is a recent model trained on synthetic instruction data enhanced with open-source code snippets. It proposes OSS-INSTRUCT which produces diverse and realistic instruction tuning data from open-source code snippets to address the biases typically found in synthetic data generated by LLMs.

\section{Conclusion}\label{sec:conclusion}
In this paper, we introduced a novel automatic LLM-tailored prompt generation method \tool for test case generation. During the optimization process, \tool generates diverse candidate prompts to facilitate the exploration of the prompt search space and induces rules from failure cases to avoid recurring errors. Additionally, \tool integrates various domain contextual knowledge for generating correct test cases in practical projects. Extensive experiments on Defects4J show that \tool outperforms existing prompt optimization methods. The replicate package of this work is publicly available at \textit{{\url{https://zenodo.org/records/14287744}}}.


\bibliographystyle{IEEEtran}
\bibliography{sigproc}
\vfill



\end{document}